\documentclass[]{aastex631}

\renewcommand{\ion}[2]{#1\,\textsc{#2}}

\newcommand{\as}{$^{\prime\prime}$}
\newcommand{\kms}{km~s$^{-1}$}

\begin{document}

\title{Center-to-limb Variations in Solar Plage using IRIS Observations}

\author[0000-0002-0786-7307]{Pradeep Kayshap}
\affiliation{School of Advanced Sciences and Languages, VIT Bhopal University, Kothrikalan, Sehore Madhya Pradesh - 46611}
\affiliation{Inter University Centre for Astronomy $\&$ Astrophysics, Post Bag - 4, Ganeshkhind, Pune-411007, India}
\author[0000-0001-9034-2925]{Peter R. Young}
\affiliation{NASA Goddard Space Flight Center, 
Greenbelt, MD 20771, USA}
\affiliation{Department of Mathematics, Physics and Electrical Engineering, Northumbria University, Newcastle upon Tyne, UK}



\begin{abstract}
The center-to-limb variations (CLV) of transition region line Gaussian fit parameters in solar plage are reported for the first time.
The Si~{\sc iv} 1402.77~{\AA} line observed by Interface Region Imaging Spectrograph (IRIS) is used.
The spectral intensity increases linearly from the disk center to the solar limb. Similarly, the non-thermal velocity also increases linearly from 23.6 km s$^{-1}$ (at the disk center) to 30.9 km s$^{-1}$ (at the solar limb). On the other hand, the Doppler velocity decreases from 8.9$\pm$1.0 km s$^{-1}$ at the disk center to  0.0 km s$^{-1}$ at the limb. This CLV pattern in solar plages is consistent with the CLV pattern reported in the quiet-Sun (QS). However, the average values of the  parameters in the solar plage are significantly higher than in the QS. The intensity and non-thermal velocity increase linearly with the magnetic field at the disk center while the Doppler velocity does not depend on the magnetic field. 

Due to the line-of-sight effect, the plasma column depth increases towards the solar limb which leads to a linear increase in the spectral intensity. Further, the increasing plasma column depth towards the solar limb adds more and more unresolved motions, and as a result, the non-thermal velocity increases from the disk center to the solar limb. In the solar plages, the higher plasma density due to the strong magnetic field leads to the higher intensity and non-thermal velocity compared to QS and coronal hole (CH). 
\end{abstract}

\keywords{Solar transition region(1532) --- Ultraviolet spectroscopy(2284) --- Center to limb observations(1972)}


\section{Introduction} \label{sec:intro}

Plages are large areas of enhanced chromospheric brightness on the solar surface that correspond to active regions. They have traditionally been identified in visible chromospheric lines, particularly, H-alpha, but are also clearly seen in full-disk images of the ultraviolet 1600\,\AA\ and 1700\,\AA\ channels of the Atmospheric Imaging Assembly \citep[AIA:][]{2012SoPh..275...17L} instrument on the Solar Dynamics Observatory (SDO). Comparison of these images with line-of-sight magnetograms from the Helioseismic and Magnetic Imager \citep[HMI:][]{2012SoPh..275..207S} on SDO readily demonstrates a connection between plage and strong magnetic fields. Plage regions correlate with the location of photospheric faculae that are collections of small magnetic flux tubes. As active regions age and their flux disperses over the solar surface, plages are replaced by enhanced network \citep{1987ARA&A..25...83Z}.\\

Plages cause the increase in total solar irradiance from minimum to maximum during the solar cycle \citep{2002A&G....43e...9S}. They remain bright into the solar transition region, and \citet{2018A&A...619A...5B} investigated spatial correlations of IRIS emission lines and AIA 1600\,\AA\ with the plage magnetic field. Correlation is greatest for the AIA channel and lowest for the \ion{Si}{iv} line, likely due to extended loop structures in the transition region line \citep{2018A&A...619A...5B}.\\

Center-to-limb variations (CLV) of Gaussian fit parameters of optically thin emission lines provide diagnostics of physical processes occurring in the solar atmosphere. Mainly, the CLV in the spectroscopic parameters (i.e., intensity, Doppler velocity, non-thermal velocity) of several chromospheric and transition-region spectral lines is studied for the quiet-Sun and coronal hole (e.g., \citealt{1984ApJ...281..870D, 1998A&A...337..287E, 1999ApJ...516..490P, 1999ApJ...522.1148P, 2002A&G....43e...9S, 2022MNRAS.511.1383R, 2023MNRAS.526..383K}). The Doppler velocity is almost zero at the solar limb, and the Doppler velocity (i.e., redshift) increases towards the disk center (e.g., \citealt{1999ApJ...522.1148P,2023MNRAS.526..383K}. Here, it should be noted that, on average, all wave and mass motion should cancel out above the solar limb, therefore, the Doppler velocity at the solar limb is zero (e.g., \cite{1976ApJ...205L.177D, 1999ApJ...522.1148P}). While the non-thermal velocity is maximum at the solar limb and linearly decreases towards the disk center (e.g., \citealt{2022MNRAS.511.1383R, 2023MNRAS.526..383K}). As we know plasma column depth increases as we move toward the solar limb, and more and more unresolved motions are added. Therefore, the non-thermal velocity is maximum at the solar limb. In \citet{2023MNRAS.526..383K}, we presented results for the \ion{Si}{iv} 1402.77\,\AA\ line for coronal hole and quiet Sun regions. In the present article, we extend this study to plage regions.

\section{Observations}

Fifteen IRIS datasets obtained over a three month period from December 2014 to March 2015 were used for analysis. 
The datasets were chosen as they each contain a plage region and spanned a complete range of heliocentric angles, $\theta$, from $\mu=\cos\theta=0$ (limb) to $\mu=1$ (disk center). In each case the same raster was used for the observation: 64 slit positions separated by 2\as\ yielding an image size of 127\as\ $\times$ 119\as, and an exposure time at each slit position of 16~s. Here, please note that the IRIS spatial resolution in the Y-direction is 0.167\as\, and the spectral resolution is 26 m\AA. The cadence in each observation is 17~s. Table~\ref{table:obs_detail} gives the start and stop times of each raster, the raster centers (Xcen, Ycen), and the ranges of $\mu$ values spanned by the rasters. 
\ion{Si}{iv} 1402.75~\AA\  was selected for the analysis as this was the line used by \citet{2023MNRAS.526..383K}, and Section~\ref{sec:fitting} describes how the spectral line fit parameters were derived.
For each dataset, co-temporal and co-aligned images from HMI and AIA were used to define the plage regions within the IRIS rasters, as described in Section~\ref{sec:selection}.

\begin{deluxetable}{cccccc}
\caption{The table provides all the necessary details about 15 plage observations that are utilized in this work.}\label{table:obs_detail}
\tablehead{
\colhead{Sr No.} & \colhead{Date} & \colhead{Start Time} & \colhead{End Time} & \colhead{Xcen, Ycen} & \colhead{$\mu$ Range}
}
\startdata
1.      & 13 Dec 2014  & 07:11:06~UT  & 07:21:34~UT & -164.46", 175.65" & 0.946{--}0.983\\ 
2.      & 14 Dec 2014  & 09:08:17~UT  & 09:26:06~UT & 81.15", 174.25"   & 0.958{--}0.993\\
3.      & 15 Dec 2014  & 06:16:17~UT  & 06:34:06~UT & 275.13", 183.87"  & 0.903{--}0.967\\
4.     & 16 Dec 2014  & 06:38:54~UT  & 06:56:44~UT  & 483.41", 183.87"  & 0.790{--}0.893\\
5.     & 18 Dec 2014  & 07:24:58~UT  & 07:42:48~UT  & 803.57", 179.86"  & 0.0383{--}0.637\\
6.     & 21 Dec 2014  & 08:35:38~UT  & 08:53:28~UT & -464.50", -236.08" & 0.0785{--}0.894\\
7.     & 12 Feb 2015  & 11:10:18~UT  & 11:28:07~UT & 670.01", 283.74" & 0.0550{--}0.745\\
8.     & 15 Feb 2015  & 06:24:18~UT  & 06:42:08~UT & 925.44", 248.63" & 0.000{--}0.414\\
9.     & 21 Feb 2015  & 05:29:20~UT  & 05:47:09~UT & -689.29", -109.94" & 0.0607{--}0.763\\
10.    & 23 Feb 2015  & 15:51:15~UT & 16:09:04~UT  & -157.89", -24.59"  & 0.970{--}0.995\\
11.    & 24 Feb 2015  & 07:09:18~UT  & 07:27:08~UT & -17.31", -28.04"   & 0.993{--}1.0\\
12.    & 26 Feb 2015  & 05:07:31~UT  & 05:25:21~UT & 370.30", -47.82" & 0.887{--}0.945\\
13.    & 01 Mar 2015  & 01:59:18~UT  & 02:17:08~UT & 842.13", -63.49" & 0.328{--}0.592\\
14.    & 03 Mar 2015  & 23:09:18~UT  & 23:27:08~UT & 475.43", 193.38" & 0.0787{--}0.893\\
15.    & 08 Mar 2015  & 09:51:18~UT  & 10:09:08~UT & 949.77", 125.25" & 0.000 -0.389\\
\enddata
\end{deluxetable}

\section{Data Analysis}

\subsection{Gaussian Fitting}\label{sec:fitting}
\begin{figure}
\mbox{
\centering
 \includegraphics[width=\textwidth, trim = 0.0cm 0.5cm 0.2cm 1.0cm,scale=1.0]{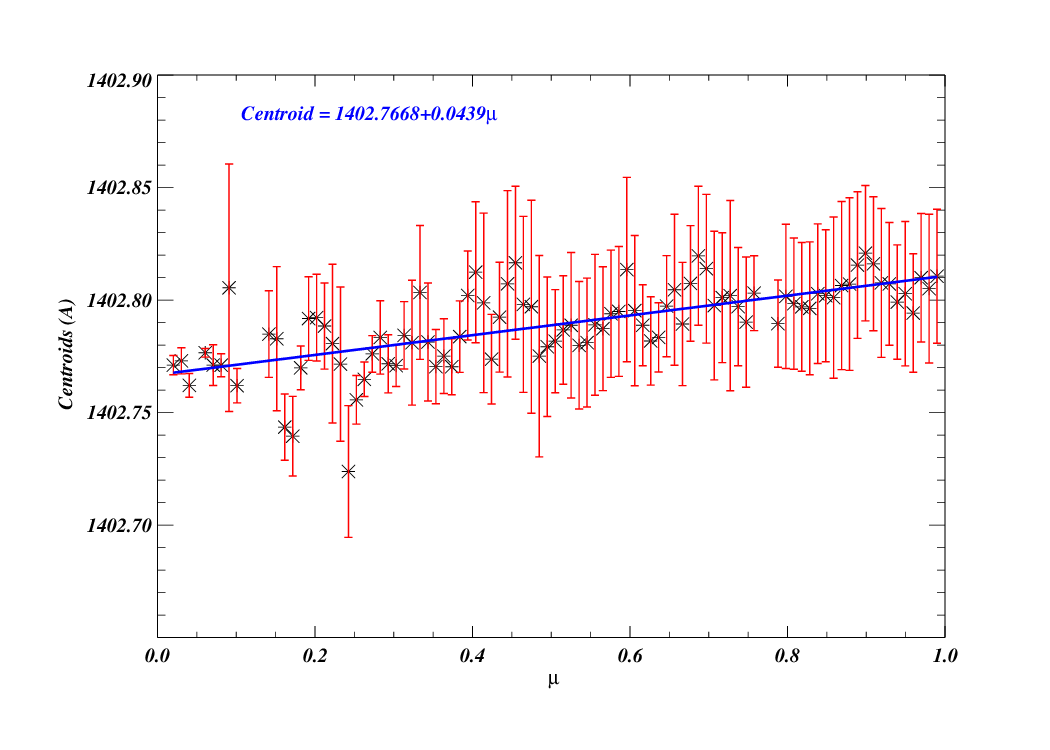}
 }
\caption{We have created 100 different $\mu$ bins from solar limb ($\mu$ = 0.0) to solar disk center ($\mu$ = 1.0), and then we estimated mean and standard errors of centroid in each $\mu$ bin. The mean centroid (from each bin) against the $\mu$ bin value is displayed in this figure. The red error bars show the corresponding standard errors. The blue line is the straight line fit, i.e., y = 1402.7668+0.0439$\mu$. Hence, the centroid at the limb is 1402.7668, i.e., the rest wavelength is 1402.77~{\AA}.}
\label{fig:centroid}
\end{figure}

Usually, the optically thin lines (such as Si{\sc iv}~1402.75~{\AA}) of solar atmosphere follow Gaussian distribution, and such lines can fit very well with the Gaussian function\footnote{https://iris.lmsal.com/itn38/analysis\_lines\_iris.html}. However, if the ions do not follow the Maxwellian distribution of velocities then the line shape deviates from the standard Gaussian shape (\cite{2017ApJ...842...19D}). Please note that the majority of the Si~{\sc iv} line has a Gaussian shape in the used observations. 
Now, by applying a Gaussian fit to the observed profiles, we estimate peak intensity (I$_{p}$), centroid ($\lambda$), and Gaussian width ($\sigma$) for each spatial pixel of the IRIS rasters. 
The $I$ map for dataset 2 is shown in Figure~\ref{fig:hmi_mag}(c). To derive Doppler velocities, it is necessary to specify a rest wavelength for the \ion{Si}{iv} line. This was obtained by first taking the complete set of centroid measurements and assigning them to 100 $\mu$ bins equally spaced between 0 and 1. Figure~\ref{fig:centroid} displays the average centroids in each bin and the errors reflect the standard deviation of the centroids within the bins. A linear fit was performed to the binned centroids (shown on Figure~\ref{fig:centroid}), which yields a wavelength of $\lambda_0=1402.7668$~\AA\ at the limb. This was used as the rest wavelength for all datasets in order to derive Doppler velocities, $v_\mathrm{D}$.

The non-thermal velocity, $\xi$, is obtained from the full-width at half-maximum (FWHM), $w$, which is related to $\sigma$ by $w=2\sqrt{2\ln 2}\,\sigma$. $w$ can be expressed as a sum in quadrature of three components:
\begin{equation}
    w^2=w_\mathrm{I}^2 + w_\mathrm{T}^2  + w_\mathrm{NT}^2 
\end{equation}
where $w_\mathrm{I}$ is the instrumental width \citep[26~m\AA:][]{DePon2014}, $w_\mathrm{T}$ is the thermal width and $w_\mathrm{NT}$ is the non-thermal width. The latter two terms are written as 
\begin{equation}
w_{T}^2 + w_{NT}^2 = 4\ln 2 \left({\lambda_{0}\over c}\right)^2\left({2kT_\mathrm{i}\over m}+\xi^2\right)
\label{eq:eq1}
\end{equation}
where $c$ is the speed of light, $k$ is the Boltzmann constant, $T_\mathrm{i}$ is the formation temperature of \ion{Si}{iv} (74\,000~K is used here), and $m$ is the mass of a silicon ion. The procedure used here is the same as that used by \cite{2023MNRAS.526..383K} where further details can be found. Maps of $v_\mathrm{D}$ and $\xi$ obtained from dataset 2 are shown in Figure~\ref{fig:hmi_mag}(d) and Figure~\ref{fig:hmi_mag}(e), respectively.

\subsection{Selection of Solar Plage Regions}\label{sec:selection}

The precise specification of the plage regions within the IRIS rasters was performed using HMI line-of-sight (LOS) magnetograms, as described below. Datasets 8 and 15 are close to the limb and there is not a good spatial correspondence between the \ion{Si}{iv} bright regions and the regions of strong LOS magnetic field. Therefore, for these datasets only, plage was defined using AIA 1600~\AA\ images.

In order to accurately identify the plage regions in the IRIS rasters, it was necessary to first create ``pseudo-rasters" from the HMI and AIA image sequences. That is, for each IRIS slit exposure, the nearest-in-time HMI/AIA image was selected and a north-south slit was extracted matching the spatial location of the IRIS slit. As the HMI and AIA plate scales differ from that of IRIS, the slit intensities were interpolated to the IRIS plate scale. The HMI/AIA slits were then assembled to create two-dimensional images that matched the IRIS raster images. The HMI/AIA pseudo-raster images were coaligned to the IRIS intensity maps using the IDL routine get$\_$correl$\_$offsets.pro.

\begin{figure}
\mbox{
\centering
 \includegraphics[trim = 0.5cm 0.0cm 0.5cm 2.0cm,scale=1.0]{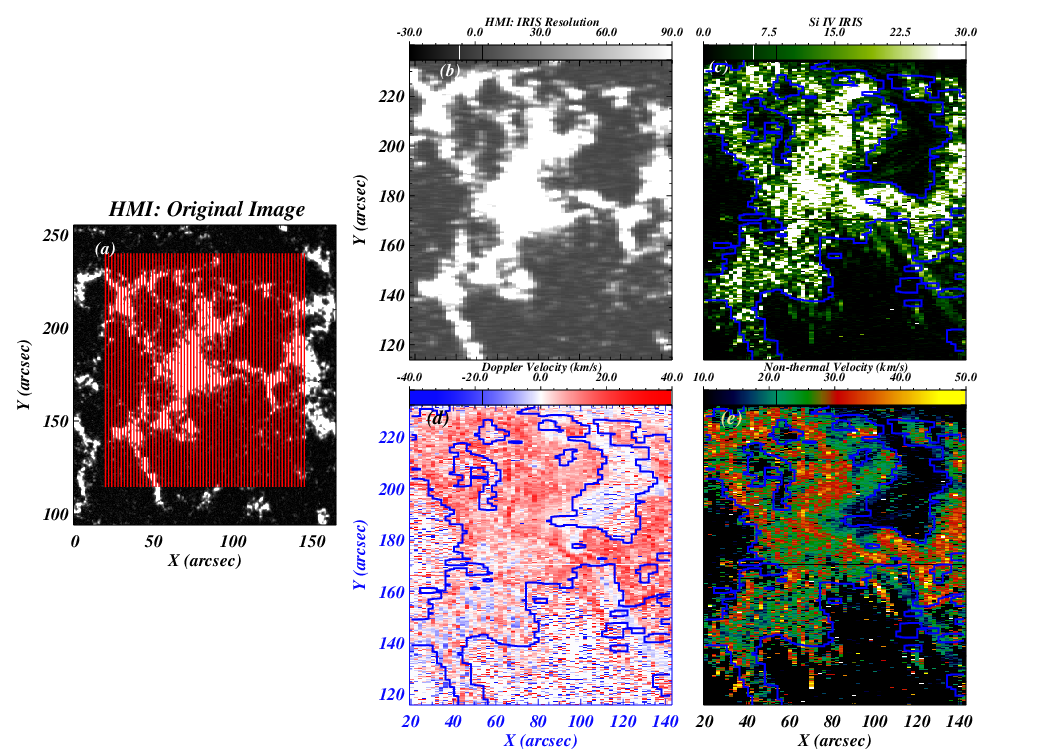}
 }
\caption{Panel (a) shows the line-of-sight (LOS) photospheric magnetogram (provided by HMI/SDO) from dataset 2 (2014 December 14). The overplotted solid red lines are the 64 slit locations used by IRIS to observe this region. Using the locations from these IRIS slits, we have produced a raster image from the HMI magnetogram corresponding to IRIS observation, which is shown in panel (b). In addition, the panels (c), (d), and (e) show the intensity, Doppler velocity, and non-thermal velocity maps. These maps are deduced by applying a Gaussian fit to the observed Si~{\sc iv} 1393.75~{\AA} line profiles at each location. The overplotted blue contours (in panels (c), (d), and (e)) outline the plage area.}
\label{fig:hmi_mag}
\end{figure}

The method for creating pseudo-raster images is illustrated in Figure~\ref{fig:hmi_mag} for dataset 2. Panel (a) shows a representative HMI image of the absolute LOS magnetic field, $|B_\mathrm{los}|$. 64 vertical red lines are over-plotted to indicate the positions of the IRIS slit. Figure~\ref{fig:hmi_mag}(b) shows the pseudo-raster image of $|B_\mathrm{los}|$, and Figure~\ref{fig:hmi_mag}(c) shows the \ion{Si}{iv} 1402.75~\AA\ intensity map for comparison.

The method for identifying plage from HMI LOS magnetograms is derived from the work of \citet{Yeo2013}. These authors defined network and faculae (equivalent to plage here) spatial pixels to have a magnetic field greater than three times the local noise level in the magnetogram. The local noise level varies approximately radially across the magnetograms, increasing from 4.9~G at disk center to 8.6~G close to the limb. These results were derived by \citet{Yeo2013} from data obtained in 2010 and we verified them using data from 2014, close to the times of the plage observations. Isolated pixels were excluded from the final plage maps, and only substantially large patches were retained. Figure~\ref{fig:hmi_mag}(c)--(e) show the plage contours (blue) overplotted on the $I$, $v_\mathrm{D}$ and $\xi$ maps for dataset 2.

For the observations close to the limb (datasets 8 and 15), pseudo-rasters were created from AIA 1600~\AA\ images and coaligned with the \ion{Si}{iv} 1402.75~\AA\ intensity images in the same manner as for the HMI data. An intensity threshold of 110~DN~s$^{-1}$~pix$^{-1}$  was applied to the 1600~\AA\ images in order to define the plage regions, and the boundaries were expanded by 5 pixels by applying the IDL dilate.pro function. Figures~\ref{fig:maks_plage_15F} and \ref{fig:maks_plage_08M} show the plage boundaries over-plotted on the IRIS line parameter maps for datasets 8 and 15.


\section{Results}

\subsection{CLV of Spectral Parameters in the Solar Plage}

The procedure described in Section~\ref{sec:fitting} was applied to all 15 IRIS datasets to yield $I$, $v_\mathrm{D}$ and $\xi$  at each spatial pixel. The $\mu$ value for each pixel was computed based on the spatial coordinates and the HMI pseudo-rasters yielded the $B_\mathrm{LOS}$ value for each pixel. The pixels were flagged as plage and non-plage based on the contours derived from the procedure described in Section~\ref{sec:selection}.

\begin{figure}
\mbox{
\centering
 \includegraphics[trim = 1.0cm 1.0cm 0.2cm 0.5cm,scale=1.2]{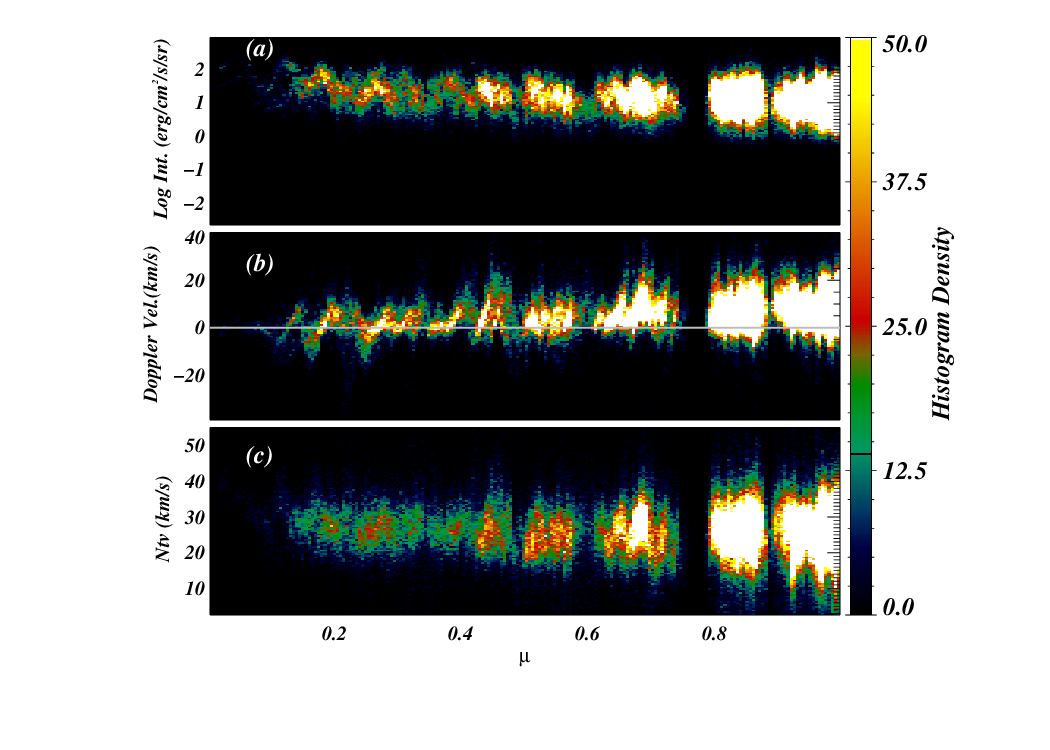}
 }
\caption{We have displayed the 2-D density distribution of the intensity (top panel), Doppler velocity (middle panel), and non-thermal velocity (bottom panel) with $\mu$. It is clearly visible that intensity and non-thermal velocity are decreasing with $\mu$ while the Doppler velocity increases from limb to disk center. The white line in the 2-D Doppler velocity distribution (panel b) shows the zero Doppler velocity.}
\label{fig:clv_plage}
\end{figure}

Taking all of the IRIS plage pixels, 2D histograms of $I$, $v_\mathrm{D}$ and $\xi$ as functions of $\mu$ were derived and are plotted in Figure~\ref{fig:clv_plage}. Both $I$ and $\xi$ generally decrease from the limb to disk center, while  $v_\mathrm{D}$ increases towards disk center. These trends are consistent with coronal hole and quiet Sun results from \citet{2023MNRAS.526..383K}.

\begin{table}
\centering
\caption{Linear fit parameters ($a,b$) and Pearson correlation coefficients (PCC) for $I$, $v_\mathrm{D}$ and $\xi$.}
\begin{tabular}{cccc}
\hline
Parameter & $a$ & $b$ & PCC \\
\hline
$\log\,I$ & 1.67$\pm$0.033  & $-0.56\pm 0.058$ & $-0.70$\\
$v_\mathrm{D}$ & 0.05$\pm$0.55 & 8.89$\pm$0.96 & $+0.68$ \\
$\xi$  & 30.84$\pm$0.49 & $-7.22\pm 0.87$ & $-0.64$ \\
\hline
\end{tabular}
\label{table:params}
\end{table}

\begin{figure}
\mbox{
\centering
 \includegraphics[trim = 2.0cm 1.5cm 2.2cm 1.0cm,scale=1.2]{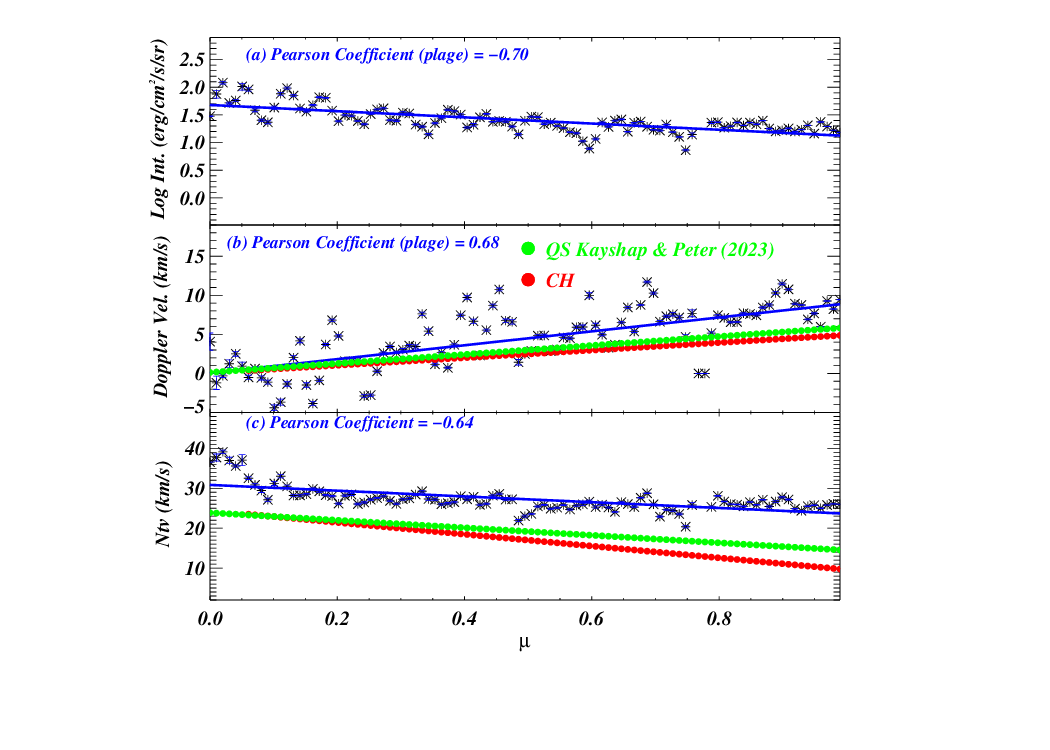}
 }
\caption{We have calculated the mean and standard error of intensity, Doppler velocity, and non-thermal velocity in different 100$\mu$ bins. These 100 $\mu$ bins cover the full $\mu$ range from 0 (disk center) to 1 (solar limb). This figure displays the variation of mean intensity (top panel), mean Doppler velocity (middle panel), and mean non-thermal velocity (bottom panel) with $\mu$. The blue error bar corresponds to the standard error. The behavior of intensity, Doppler velocity, and non-thermal velocity are similar as we have seen in Figure~\ref{fig:clv_plage}. The blue line in each panel corresponds to the linear fit on the respective parameter of solar plage. Pearson's coefficients for solar plages are also mentioned in each panel. In addition, the green and red dotted lines are the linear fit of the Doppler velocities derived from QS and CH, respectively (panel b). Similarly, we have also displayed non-thermal velocities from QS (green dotted curve) and CH (red dotted curve) in panel (c). The Doppler and non-thermal velocities are taken from \citealt{2023MNRAS.526..383K}.}
\label{fig:clv_plage_avg}
\end{figure}

An alternative method for displaying the plage data is shown in Figure~\ref{fig:clv_plage_avg}. The data have been separated into 100 $\mu$ bins from 0 to 1 in equal intervals of 0.01. For each spatial pixel in a bin, $I$, $v_\mathrm{D}$ and $\xi$ are each averaged and standard errors computed. Figure~\ref{fig:clv_plage_avg} shows the average values and errors. Linear fits to the three parameters are shown on each plot, and the Pearson correlation coefficients (PCC) between the parameters and $\mu$ are given. For $v_\mathrm{D}$ and $\xi$ the linear fits obtained by \citet{2023MNRAS.526..383K} for quiet Sun and coronal hole are over-plotted for comparison. The linear fit parameters, defined as $a+b\mu$, are given in Table~\ref{table:params} with the PCC values.

The Doppler velocity at disk center determined from the linear fit is 8.9~\kms, significantly higher than the values found by \citet{2023MNRAS.526..383K} for coronal hole and quiet Sun. In addition, the non-thermal velocity is systematically higher in plage by around 5--10~\kms\ for all $\mu$ values.

\subsection{Variations of Spectral Parameters with the Magnetic Field}

\begin{figure}
\mbox{
\centering
 \includegraphics[trim = 1.0cm 0.0cm 0.2cm 1.0cm,scale=1.2]{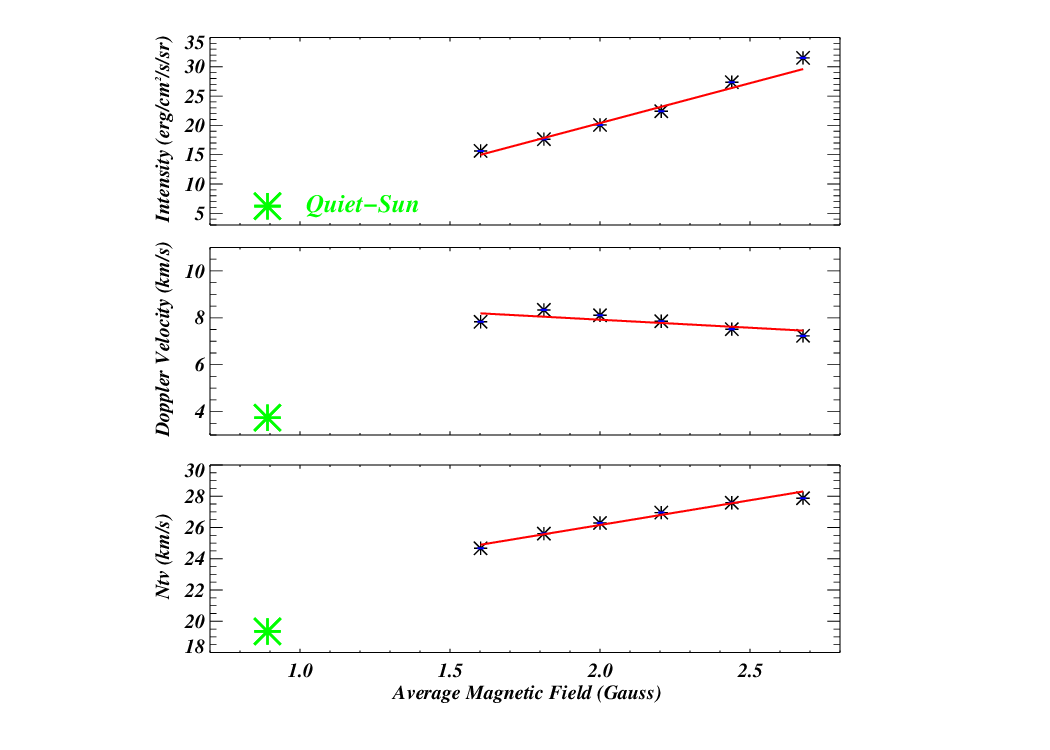}
 }
\caption{The average of intensity (top panel), Doppler velocity (middle panel), and non-thermal velocity (bottom panel) vs magnetic field. We have averaged over all $\mu$ values for each magnetic field bin. The red line in each panel is the linear fit on the corresponding parameter. In addition, we have also displayed the average values of intensity (panel a), Doppler velocity (panel b), and non-thermal velocity (c) by the green asterisk signs from the QS.}
\label{fig:mag_para}
\end{figure}

Plage is defined based on the strength of the magnetic field. In this section we investigate whether the average line fit parameters vary with the strength of $|B_\mathrm{LOS}|$ within the plage. For this purpose, we consider two IRIS observations close to disk center: datasets 2 and 3 (Table~\ref{table:obs_detail}). The magnetic field within the plage for these datasets is divided into six bins: 0{--}50, 50{--}80, 80{--}120, 120{--}200, 200{--}350, and 350{--}600~G. The parameters $I$, $v_\mathrm{D}$ and $\xi$ are averaged for each bin and the results are plotted in Figure~\ref{fig:mag_para} against $<|B_\mathrm{LOS}|>$, the mean value of $|B_\mathrm{LOS}|$ within the bins. Both $I$ and $\xi$ show increases with magnetic field strength, but  $v_\mathrm{D}$  shows no dependence on $B$. The variations of the parameters with $\log\,(<|B_\mathrm{LOS}|>)$ are approximately linear, and fits are over-plotted on Figure~\ref{fig:mag_para}

The green stars on the Panels of Figure~\ref{fig:mag_para} are derived from the non-plage regions of the dataset 2 and 3 rasters. For each of the three spectral line parameters, the non-plage values are smaller than the plage values. Whereas the QS intensity is consistent with a linear extrapolation of the plage trend with magnetic field, the $\xi$ and $v_\mathrm{D}$ QS values are not. The difference for $v_\mathrm{D}$ is particularly striking and suggests that different physics applies in QS regions compared to the stronger magnetic field plage regions.


\section{Discussion and Conclusion} \label{sec:cite}

To the best of the authors' knowledge, the CLV of a transition region line's Gaussian line fit parameters has not previously been reported for plage. Here we have used IRIS measurements of the \ion{Si}{iv} 1402.77\,\AA\ line, which is formed at a temperature of 80~kK. Fifteen IRIS datasets were chosen to span the range $\mu=0$ (solar limb) to $\mu=1$ (disk center). In addition, for the plage datasets close to disk center we have investigated how the line parameters vary with the strength of the magnetic field. The main findings are summarized below, where they are compared with the quiet Sun (QS) and coronal hole (CH) results of \citet{2023MNRAS.526..383K}.
%
\begin{itemize}
    \item The intensity, $I$, has a maximum at the solar limb and $\log\,I$ decreases linearly with $\mu$ towards the disk center (Figure~\ref{fig:clv_plage_avg}a). 
    \item The Doppler velocity, $v_\mathrm{D}$, is defined to be zero at the limb and increases linearly with $\mu$ to a value of $8.9\pm 1.0$~\kms\ at disk center. The trend matches that for QS and CH  but the disk center value is significantly larger than the values of 5.7 and 4.9~\kms\ found for these regions \citep{2023MNRAS.526..383K}. The Doppler velocity signifies the motions that exist in the emitting plasma. Solar plage has a stronger and more complex magnetic field configuration compared to the strength and configuration of the magnetic field of QS and CH. And, the strong and complex magnetic field leads to more dynamics in solar plage compared to QS and CH. This is the most probable reason behind the higher Doppler velocity in solar plage compared to QS and CH.
    \item The non-thermal velocity, $\xi$, has a  maximum value of 30.9~\kms\ at the limb and  decreases linearly with $\mu$ towards the disk center where it has a value of 23.6~\kms. The behaviour is again consistent with QS and CH, but the values are systematically higher in the plage for all $\mu$ values (Figure~\ref{fig:clv_plage_avg}c). At disk center the plage velocity is 9 and 14~\kms\ higher than for QS and CH, respectively. 
    \item For plage regions close to disk center, $I$ and $\xi$ were found to increase with the strength of the magnetic field in the plage; $v_\mathrm{D}$ is independent of the magnetic field strength.
\end{itemize}


The line intensity increases towards the solar limb due to an increase in the plasma column depth because of the line-of-sight effect. In general, the intensity in plage for the full $\mu$ range is higher compared to the QS and CH (\citealt{2023MNRAS.526..383K}) which is probably the result of higher density in the plage. In addition,  the intensity of the solar plage increases with the magnetic field at $\mu$ = 1.0 (near the disk center). This is likely due to plasma density increasing with the magnetic field which leads to the higher intensity. Near the disk center, the Doppler velocity of the solar plage  is almost double the Doppler velocity of QS and CH  (Figure~\ref{fig:clv_plage_avg}b). 
The non-thermal velocity is the result of unresolved motions in the in-situ plasma, therefore, the plage regions have more unresolved plasma motions compared to the QS and CH for all $\mu$ values. In addition,  the non-thermal velocity increases with the magnetic field (panel c; Figure~\ref{fig:mag_para}).
\begin{acknowledgments}
PRY acknowledges support from the Goddard Space Flight Center (GSFC) Internal Scientist Funding Model competitive work package programme. IRIS is a NASA Small Explorer mission developed and operated by LMSAL with mission operations executed at NASA Ames Research Center and major contributions to downlink communications funded by ESA and the Norwegian Space Centre. We also acknowledge the data from SDO/AIA.
\end{acknowledgments}

\appendix 
\label{sect:appendix}
\counterwithin{figure}{section}
\setcounter{figure}{0}
\section{Plages Selection in Various Observations} \label{sect:appendix_obs}

\begin{figure}
\mbox{
\centering
 \includegraphics[trim = 1.0cm 0.0cm 2.0cm 1.0cm,scale=1.2]{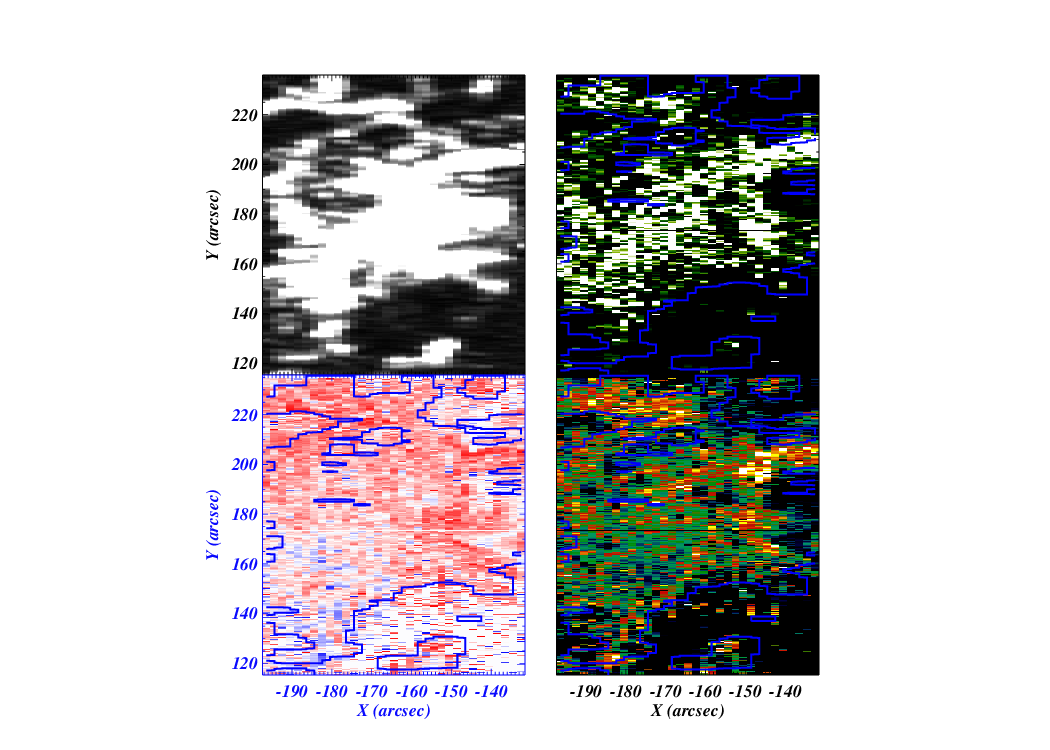}
 }
\caption{Same as panels (b), (c), (d), and (e) of Figure~\ref{fig:hmi_mag} but for 13$^{th}$ December, 2014 observation.}

\label{fig:maks_plage_13D}
\end{figure}   

\begin{figure}
\mbox{
\centering
 \includegraphics[trim = 1.0cm 0.0cm 2.0cm 1.0cm,scale=1.2]{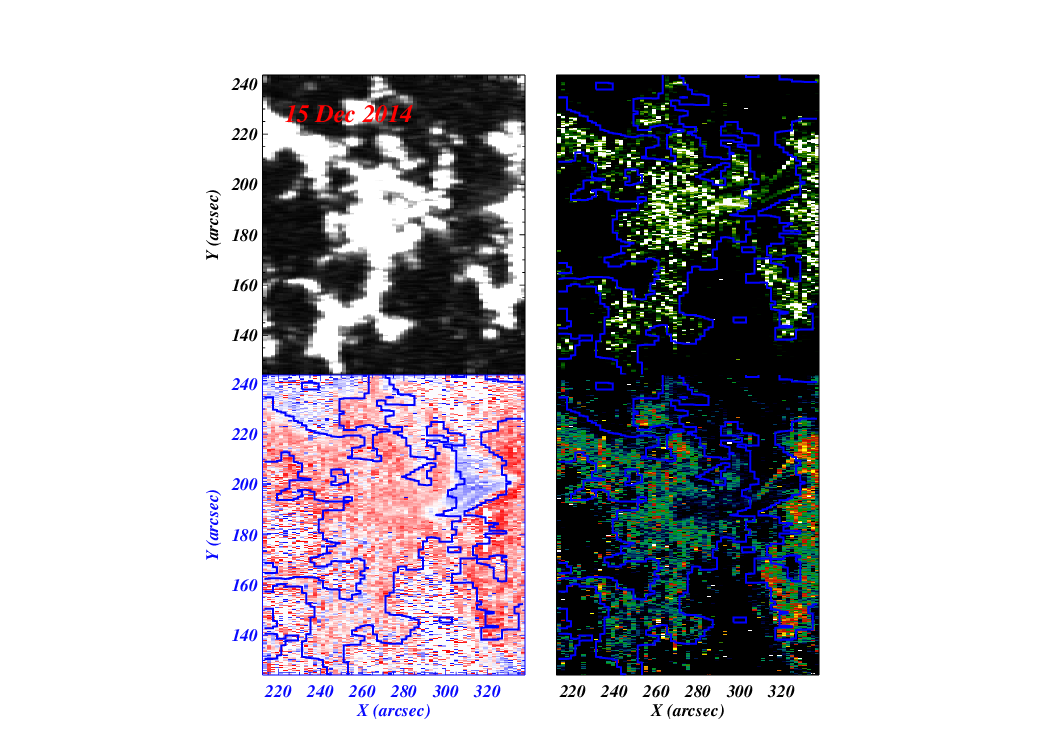}
 }
\caption{Same as panels (b), (c), (d), and (e) of Figure~\ref{fig:hmi_mag} but for 15$^{th}$ December, 2014 observation.}

\label{fig:maks_plage_15D}
\end{figure}   

\begin{figure}
\mbox{
\centering
 \includegraphics[trim = 1.0cm 0.0cm 2.0cm 1.0cm,scale=1.2]{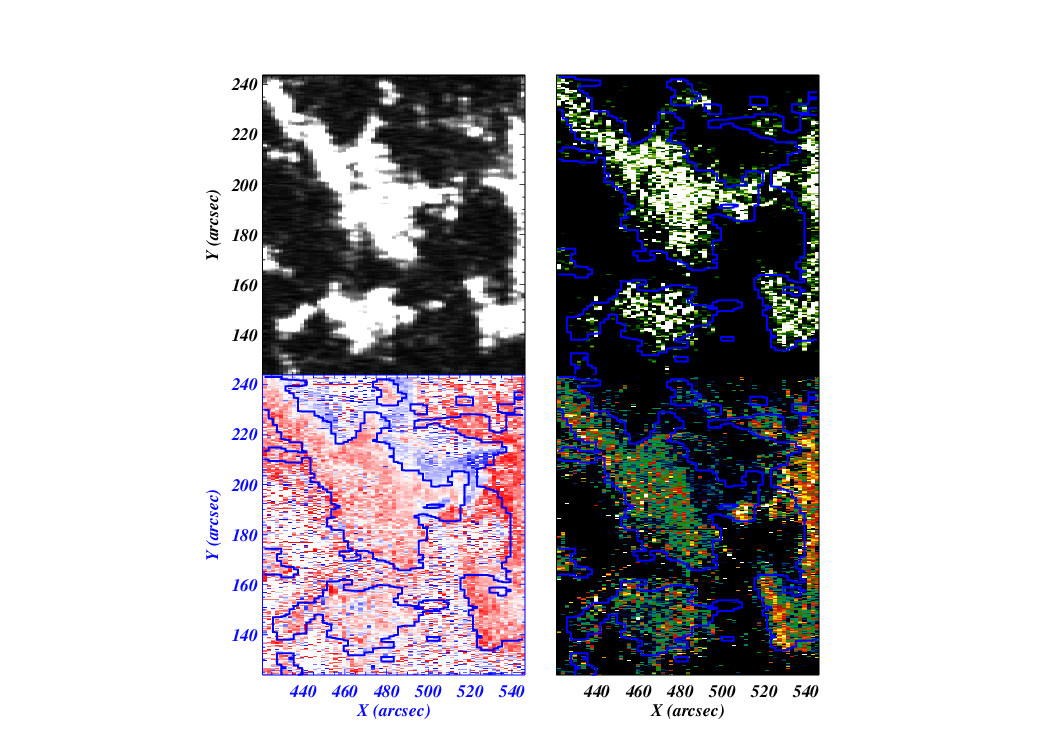}
 }
\caption{Same as panels (b), (c), (d), and (e) of Figure~\ref{fig:hmi_mag} but for 16$^{th}$ December, 2014 observation.}

\label{fig:maks_plage_16D}
\end{figure}   

\begin{figure}
\mbox{
\centering
 \includegraphics[trim = 1.0cm 0.0cm 2.0cm 1.0cm,scale=1.2]{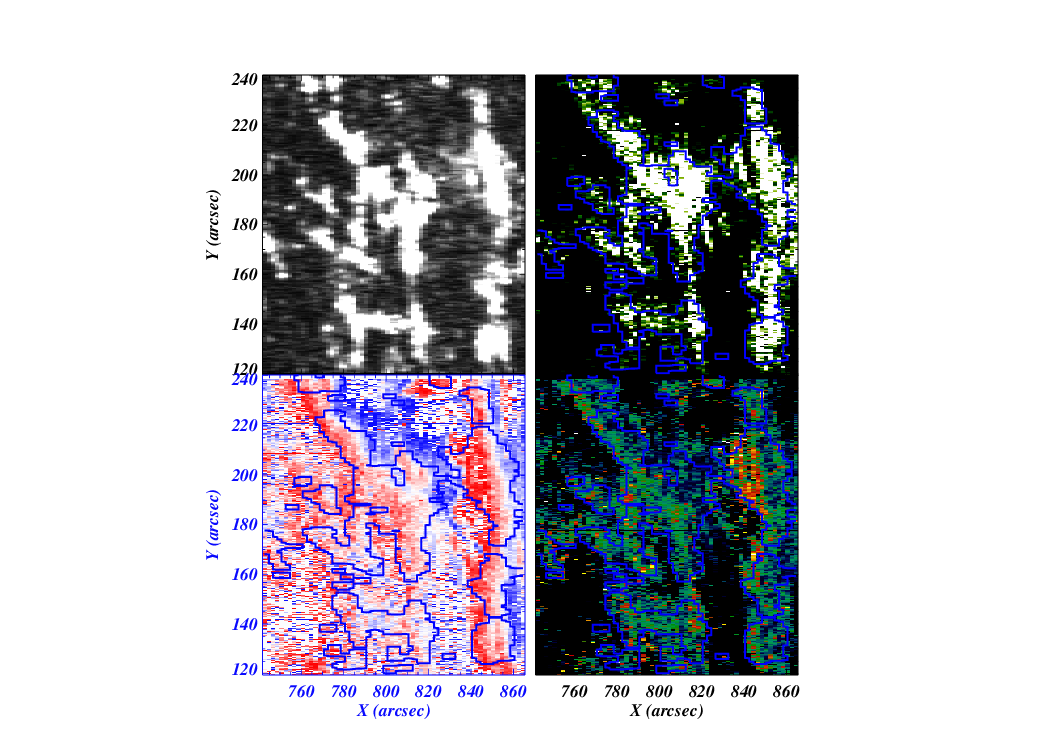}
 }
\caption{Same as panels (b), (c), (d), and (e) of Figure~\ref{fig:hmi_mag} but for 18$^{th}$ December, 2014 observation.}

\label{fig:maks_plage_18D}
\end{figure}   

\begin{figure}
\mbox{
\centering
 \includegraphics[trim = 1.0cm 0.0cm 2.0cm 1.0cm,scale=1.2]{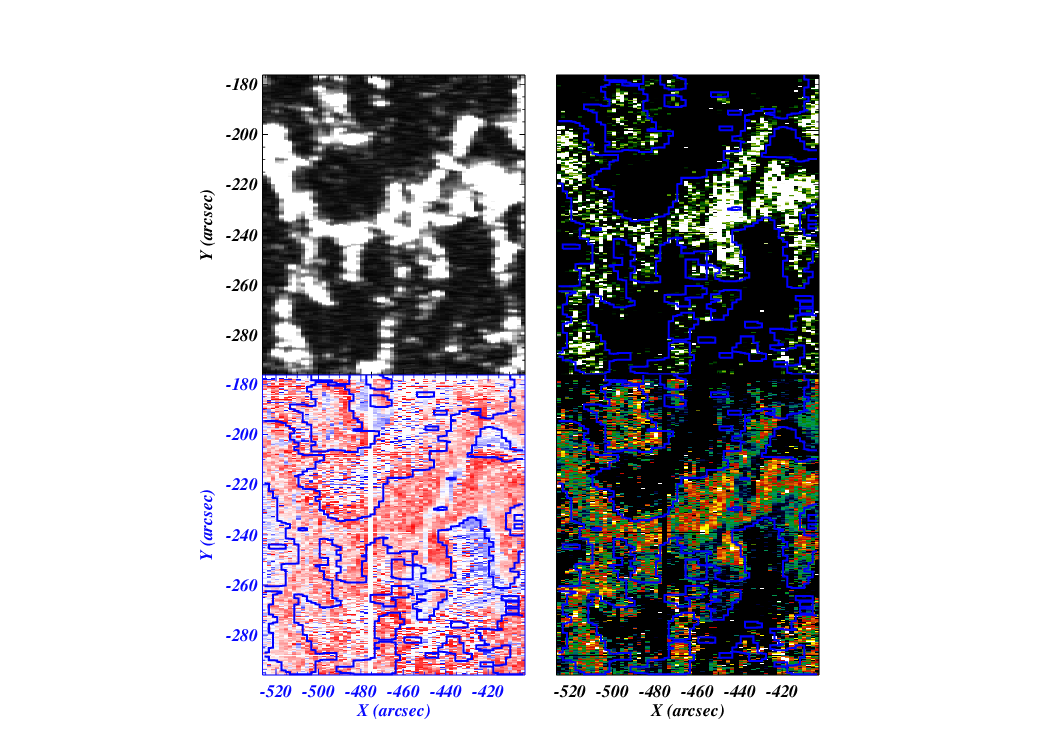}
 }
\caption{Same as panels (b), (c), (d), and (e) of Figure~\ref{fig:hmi_mag} but for 21$^{th}$ December, 2014 observation.}

\label{fig:maks_plage_21D}
\end{figure}   

\begin{figure}
\mbox{
\centering
 \includegraphics[trim = 1.0cm 0.0cm 2.0cm 1.0cm,scale=1.2]{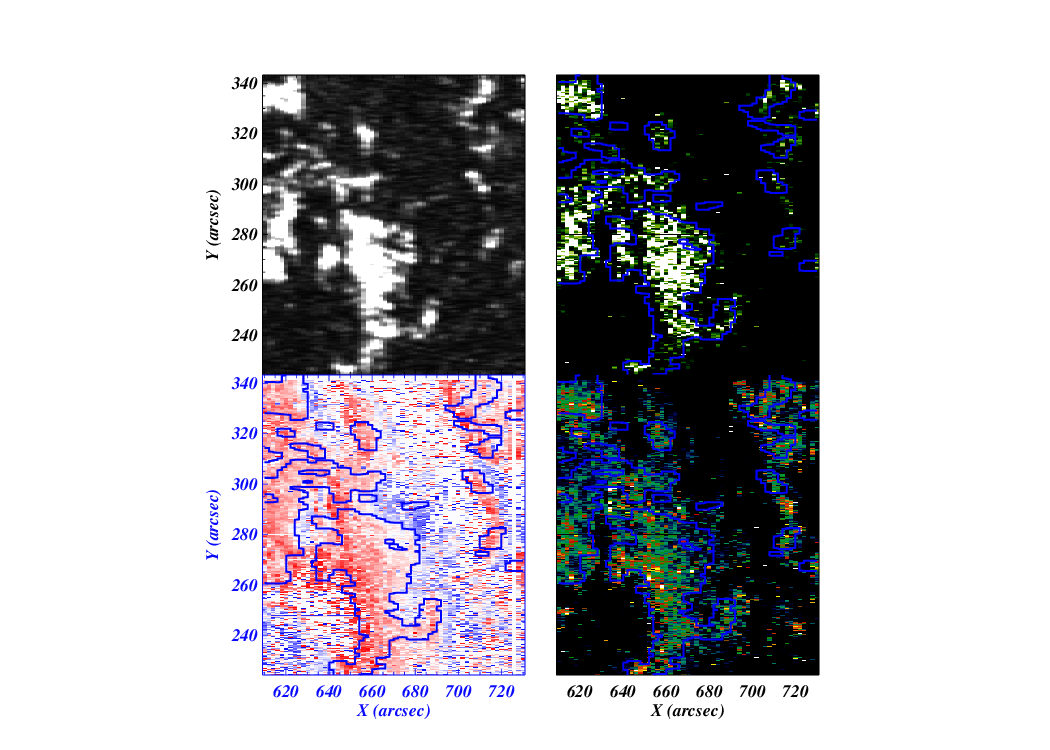}
 }
\caption{Same as panels (b), (c), (d), and (e) of Figure~\ref{fig:hmi_mag} but for 12$^{th}$ February, 2015 observation.}

\label{fig:maks_plage_12F}
\end{figure}   

\begin{figure}
\mbox{
\centering
 \includegraphics[trim = 1.0cm 0.0cm 2.0cm 1.0cm,scale=1.2]{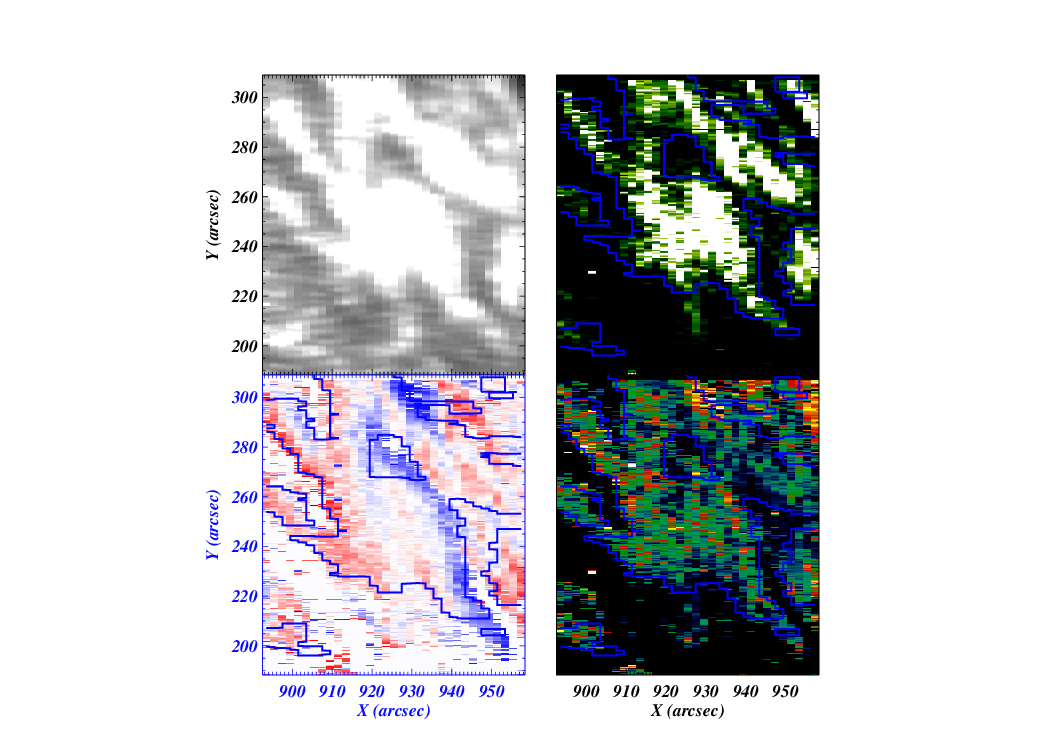}
 }
\caption{Same as panels (b), (c), (d), and (e) of Figure~\ref{fig:hmi_mag} but for 15$^{th}$ February, 2015 observation. However, it should be noted that we have considered AIA~1600~{\AA} observations (instead of HMI LOS magnetogram) to locate the plage regions. Hence, in this figure, the top-left panel is AIA~1600~{\AA} image, i.e., the AIA 1600~{\AA} raster image is created in the same way as we created magnetogram raster image, see section~\ref{sec:selection}.}

\label{fig:maks_plage_15F}
\end{figure}   

\begin{figure}
\mbox{
\centering
 \includegraphics[trim = 1.0cm 0.0cm 2.0cm 1.0cm,scale=1.2]{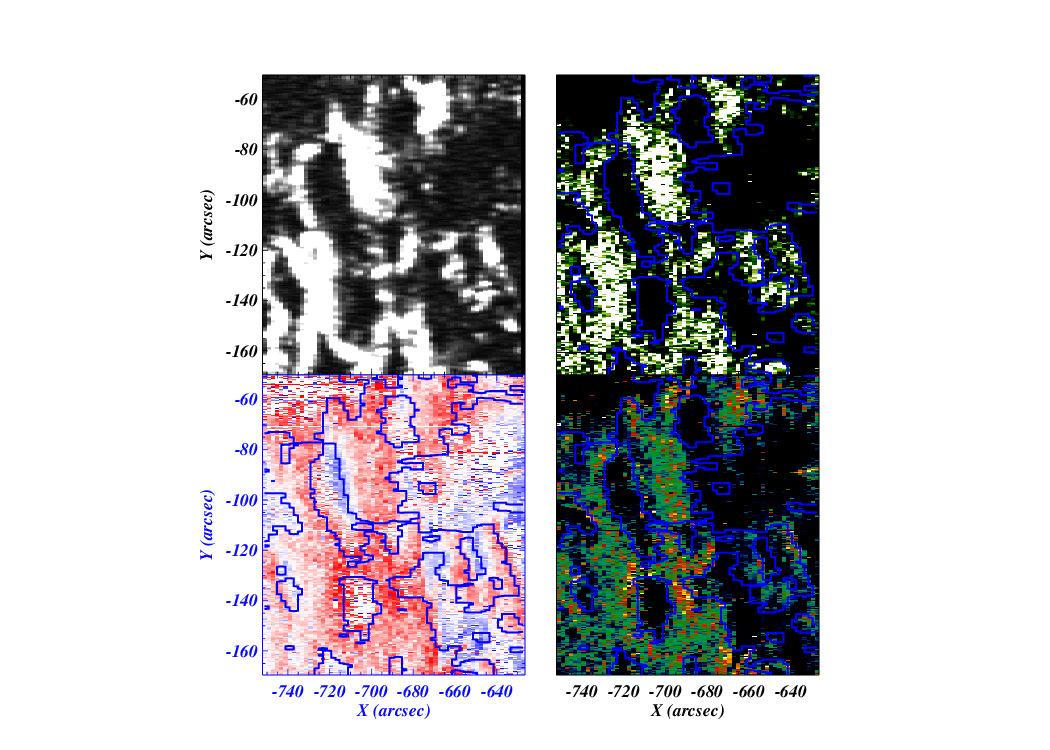}
 }
\caption{Same as panels (b), (c), (d), and (e) of Figure~\ref{fig:hmi_mag} but 21$^{th}$ February, 2015 observation.}

\label{fig:maks_plage_21F}
\end{figure}   

\begin{figure}
\mbox{
\centering
 \includegraphics[trim = 1.0cm 0.0cm 2.0cm 1.0cm,scale=1.2]{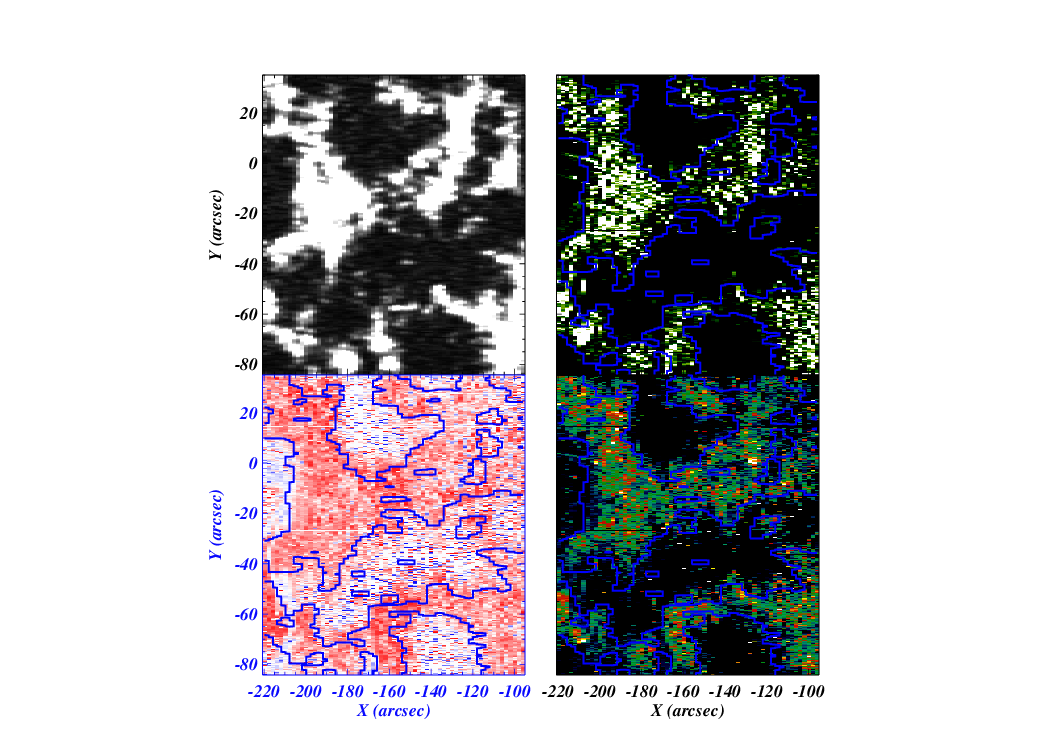}
 }
\caption{Same as panels (b), (c), (d), and (e) of Figure~\ref{fig:hmi_mag} but for 23$^{th}$ February, 2015 observation.}

\label{fig:maks_plage_23F}
\end{figure}   

\begin{figure}
\mbox{
\centering
 \includegraphics[trim = 1.0cm 0.0cm 2.0cm 1.0cm,scale=1.2]{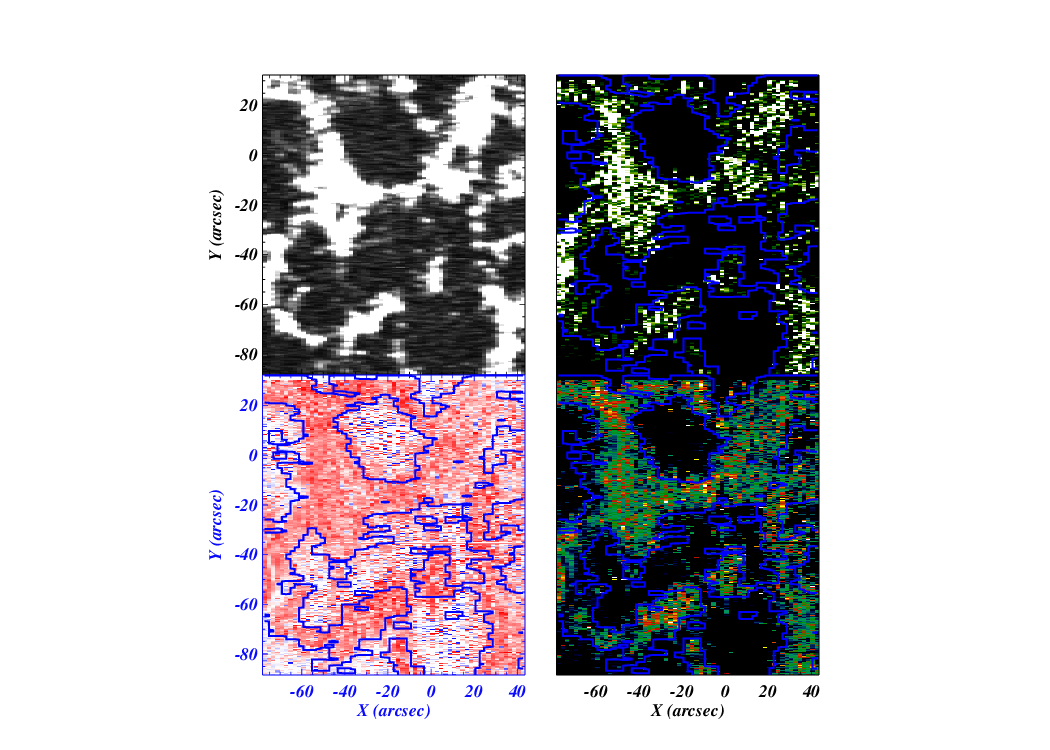}
 }
\caption{Same as panels (b), (c), (d), and (e) of Figure~\ref{fig:hmi_mag} but for 24$^{th}$ February 2015 observation.}

\label{fig:maks_plage_24F}
\end{figure}   

\begin{figure}
\mbox{
\centering
 \includegraphics[trim = 1.0cm 0.0cm 2.0cm 1.0cm,scale=1.2]{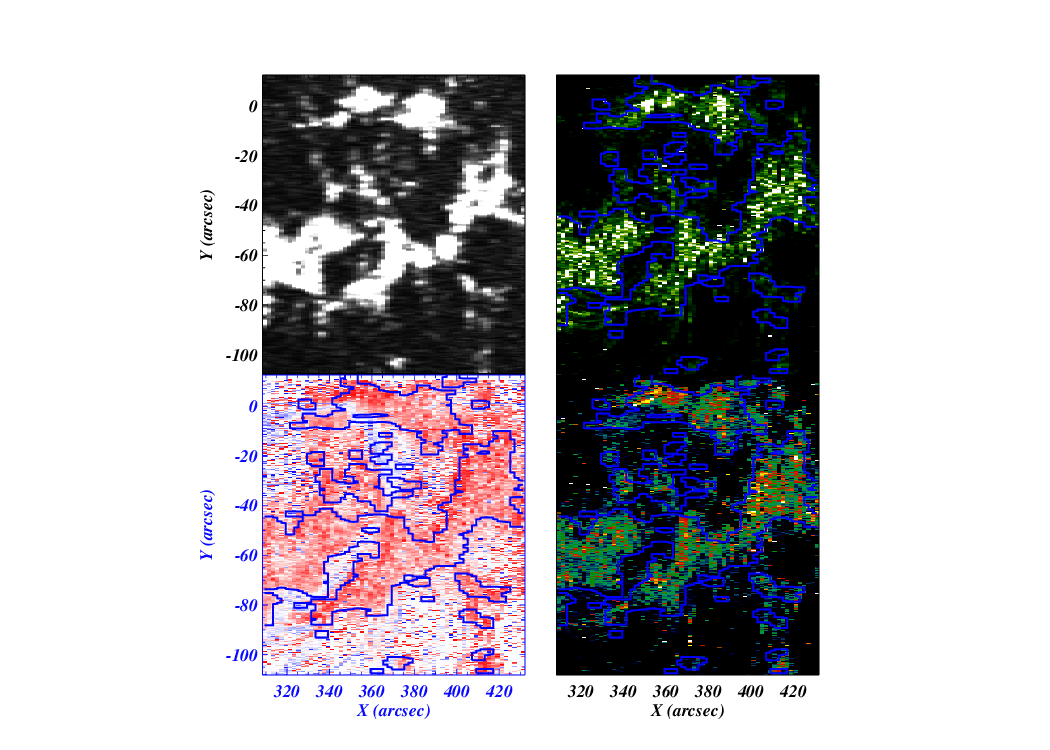}
 }
\caption{Same as panels (b), (c), (d), and (e) of Figure~\ref{fig:hmi_mag} but for 26$^{th}$ February, 2015 observation.}

\label{fig:maks_plage_26F}
\end{figure}   

\begin{figure}
\mbox{
\centering
 \includegraphics[trim = 1.0cm 0.0cm 2.0cm 1.0cm,scale=1.2]{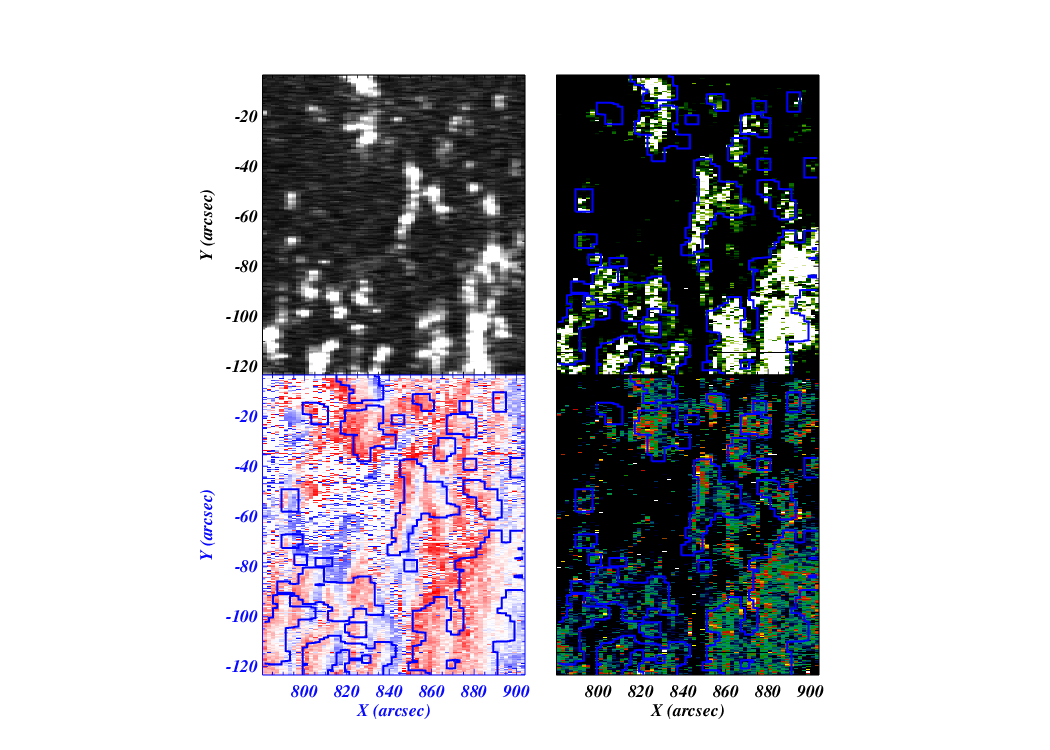}
 }
\caption{Same as panels (b), (c), (d), and (e) of Figure~\ref{fig:hmi_mag} but for 01$^{st}$ March, 2015 observation.}

\label{fig:maks_plage_01M}
\end{figure}   

\begin{figure}
\mbox{
\centering
 \includegraphics[trim = 1.0cm 0.0cm 2.0cm 1.0cm,scale=1.2]{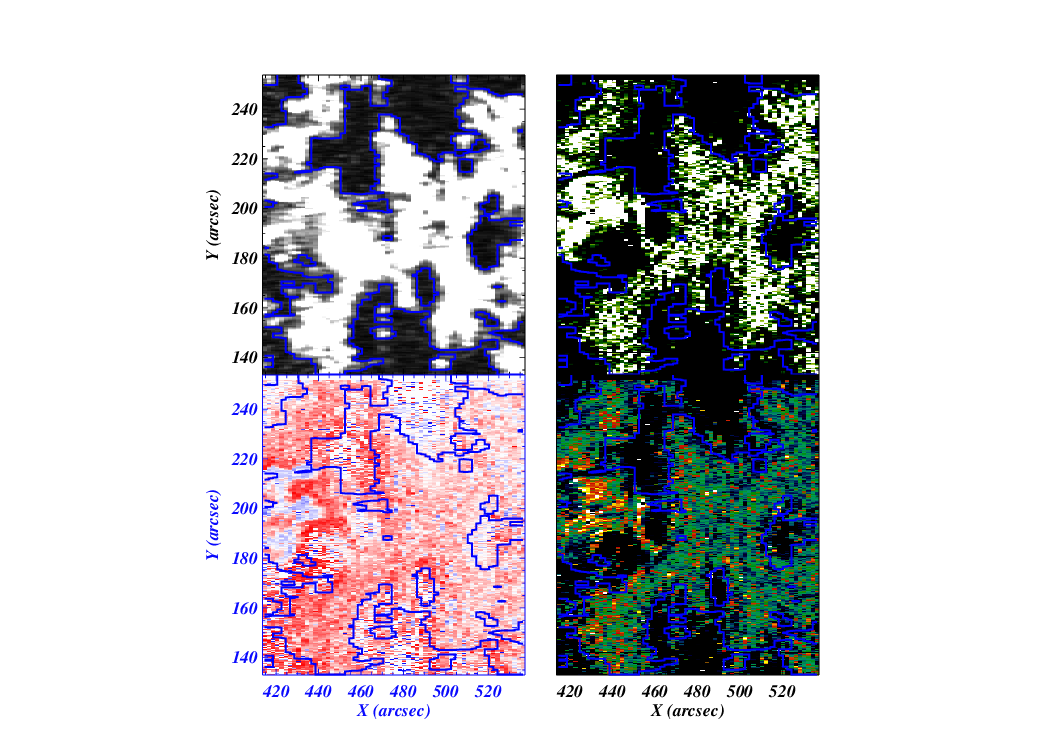}
 }
\caption{Same as panels (b), (c), (d), and (e) of Figure~\ref{fig:hmi_mag} but for 03$^{rd}$ March, 2015 observation.}

\label{fig:maks_plage_03M}
\end{figure}   

\begin{figure}
\mbox{
\centering
 \includegraphics[trim = 1.0cm 0.0cm 2.0cm 1.0cm,scale=1.2]{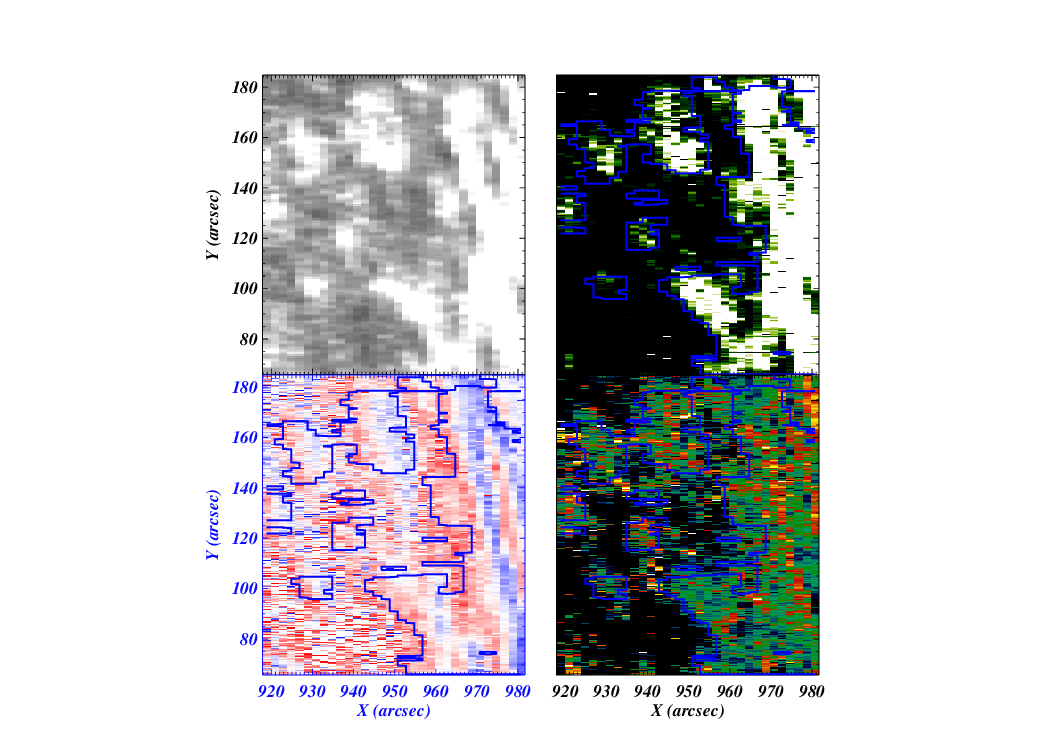}
 }
\caption{Same as Figure~\ref{fig:maks_plage_15F} but for 08$^{th}$ March, 2015 observation.}

\label{fig:maks_plage_08M}
\end{figure}   


\bibliography{sample631}{}

\begin{thebibliography}{}

\bibitem[ Barczynski et al. (2018)]{2018A&A...619A...5B} Barczynski, K., Peter, H., Chitta, L.~P., Solanki, S.~K.,2018, \aap, 619, A5,doi:10.1051/0004-6361/201731650


\bibitem[De Pontieu et al. (2014)]{DePon2014} De Pontieu, B., Title, A.~M., Lemen, J.~R. et al.\, 2014,\solphys, 289, 2733,doi:10.1007/s11207-014-0485-y
      
\bibitem[Dere et al. (1984)]{1984ApJ...281..870D} Dere, K.~P., Bartoe, J. -D.~F., Brueckner, G.~E., 1984, \apj,281,870,doi:10.1086/162167

\bibitem[Doscheck et al. (1976)]{1976ApJ...205L.177D} Doschek, G.~A., Feldman, U., Bohlin, J.~D., 1976,\apjl,205,L177,doi:10.1086/182118
     
\bibitem[Dudik et al. (2017)]{2017ApJ...842...19D} Dudik, Jaroslav, Polito, Vanessa, Dzifcakova, Elena, Del Zanna, Giulio, Testa, Paola, 2017, \apj, 842, 19, doi:10.3847/1538-4357/aa71a8

\bibitem[Erdelyi et al. (1998)]{1998A&A...337..287E} Erdelyi, R., Doyle, J.~G., Perez, M.~E., Wilhelm, K.,1998,\aap,337,287

\bibitem[Kayshap $\&$ Young (2023)]{2023MNRAS.526..383K} Kayshap, Pradeep, Young, Peter R., 2023, \mnras,526,383,doi:10.1093/mnras/stad2761
     
\bibitem[Lemen et al.(2012)]{2012SoPh..275...17L} Lemen, J.~R., Title, A.~M., Akin, D.~J., et al.\ 2012, \solphys, 275, 17. doi:10.1007/s11207-011-9776-8

\bibitem[Peter (1999)]{1999ApJ...516..490P} Peter, H., 1999, \apj, 516, 490,doi:10.1086/307102

\bibitem[Peter $\&$ Judge (1999)]{1999ApJ...522.1148P} Peter, H.,Judge, P.~G., 1999, \apj,522,1148,doi:10.1086/307672
     

\bibitem[Rao et al. (2022)]{2022MNRAS.511.1383R} Rao, Yamini K., Del Zanna, Giulio, Mason, Helen E., 2022, \mnras,511,1383,doi:10.1093/mnras/stac128


\bibitem[Scherrer et al. (2012)]{2012SoPh..275..207S} Scherrer, P.~H., Schou, J., Bush, R.~I., Kosovichev, A.~G., Bogart, R.~S., et al.\ 2012, \solphys,275,207,doi:10.1007/s11207-011-9834-2
\bibitem[Solanki (2002)]{2002A&G....43e...9S} Solanki, Sami K., 2002, Astronomy and Geophysics, 43, 5.09-5.13,doi:10.1046/j.1468-4004.2002.43509.x     

\bibitem[Yeo et al. (2013)]{Yeo2013} Yeo, K.~L., Solanki, S.~K., Krivova, N.~A., 2013, \aap, 550,A95,doi:10.1051/0004-6361/201220682
   
\bibitem[Zwaan (1987)]{1987ARA&A..25...83Z} Zwaan, Cornelis,\araa,25,83,1987,doi:10.1146/annurev.aa.25.090187.000503
       

\end{thebibliography}
\bibliographystyle{aasjournal}



\end{document}